\documentclass[superscriptaddress,prx,twocolumn]{revtex4-2}
\usepackage{graphicx}
\usepackage{amsmath,amssymb,physics,dsfont}
\usepackage{color}
\usepackage{hyperref}
\usepackage{microtype}
\usepackage{braket}
\usepackage{defs-private}

\usepackage[dvipsnames]{xcolor}
\usepackage{hyperref}% add hypertext capabilities
\hypersetup{
    colorlinks=true,
    linkcolor=BurntOrange,
    urlcolor=BrickRed,
    citecolor=Blue,
    pdftitle={Ab-initio Simulations of Coherent Phonon-Induced Pumping of Carriers in Zirconium Pentatelluride},
    pdfauthor={Tao Jiang, Peter P. Orth, Liang Luo, Lin-Lin Wang, Feng Zhang, Cai-Zhuang Wang, Jin Zhao, Kai-Ming Ho, Jigang Wang, Yong-Xin Yao}
}
\begin{document}

\title{\textit{Ab-initio} Simulations of Coherent Phonon-Induced Pumping of Carriers in Zirconium Pentatelluride}

\author{Tao Jiang}
\affiliation{Ames National Laboratory, U.S. Department of Energy, Ames, Iowa 50011, USA}

\author{Peter P. Orth}
\affiliation{Ames National Laboratory, U.S. Department of Energy, Ames, Iowa 50011, USA}
\affiliation{Department of Physics and Astronomy, Iowa State University, Ames, Iowa 50011, USA}
\affiliation{Department of Physics, Saarland University, 66123 Saarbr\"ucken, Germany}

\author{Liang Luo}
\affiliation{Ames National Laboratory, U.S. Department of Energy, Ames, Iowa 50011, USA}

\author{Lin-Lin Wang}
\affiliation{Ames National Laboratory, U.S. Department of Energy, Ames, Iowa 50011, USA}

\author{Feng Zhang}
\affiliation{Ames National Laboratory, U.S. Department of Energy, Ames, Iowa 50011, USA}
\affiliation{Department of Physics and Astronomy, Iowa State University, Ames, Iowa 50011, USA}

\author{Cai-Zhuang Wang}
\affiliation{Ames National Laboratory, U.S. Department of Energy, Ames, Iowa 50011, USA}
\affiliation{Department of Physics and Astronomy, Iowa State University, Ames, Iowa 50011, USA}

\author{Jin Zhao}
\affiliation{ICQD/Hefei National Laboratory for Physical Sciences at Microscale, and Key Laboratory of Strongly-Coupled Quantum Matter Physics, Chinese Academy of Sciences, and Department of Physics, University of Science and Technology of China, Hefei, China}
\affiliation{Department of Physics and Astronomy, University of Pittsburgh, Pittsburgh, Pennsylvania, USA}
\affiliation{Synergetic Innovation Center of Quantum Information \& Quantum Physics, University of Science and Technology of China, Hefei, China}

\author{Kai-Ming Ho}
\affiliation{Ames National Laboratory, U.S. Department of Energy, Ames, Iowa 50011, USA}
\affiliation{Department of Physics and Astronomy, Iowa State University, Ames, Iowa 50011, USA}

\author{Jigang Wang}
\affiliation{Ames National Laboratory, U.S. Department of Energy, Ames, Iowa 50011, USA}
\affiliation{Department of Physics and Astronomy, Iowa State University, Ames, Iowa 50011, USA}

\author{Yong-Xin Yao}
\email{ykent@iastate.edu}
\affiliation{Ames National Laboratory, U.S. Department of Energy, Ames, Iowa 50011, USA}
\affiliation{Department of Physics and Astronomy, Iowa State University, Ames, Iowa 50011, USA}

\begin{abstract}

\begin{center}
\textbf{Abstract}
\end{center}

Laser-driven coherent phonons can act as modulated strain fields and modify the adiabatic ground state topology of quantum materials. Here we use time-dependent first-principles and effective model calculations to simulate the effect of the coherent phonon induced by strong terahertz electric field on electronic carriers in the topological insulator ZrTe$_5$. We show that a coherent $A_\text{1g}$ Raman mode modulation can effectively pump carriers across the band gap, even though the phonon energy is about an order of magnitude smaller than the equilibrium band gap. We reveal the microscopic mechanism of this effect which occurs via Landau-Zener-St\"uckelberg tunneling of Bloch electrons in a narrow region in the Brillouin zone center where the transient energy gap closes when the system switches from strong to weak topological insulator. The quantum dynamics simulation results are in excellent agreement with recent pump-probe experiments in ZrTe$_5$ at low temperature.
\end{abstract}

\maketitle

\section{Introduction}
\label{sec:introduction}
Coherent phonons that are excited by laser pulses in the THz or mid-infrared frequency range~\cite{forst2011nonlinear, merlin1997generating, dhar1994time} can provide nonthermal pathways for the dynamical control of quantum phases of condensed matter~\cite{basov2017towards, mankowsky2016non, de2021Nonthermal, disa2021engineering}. Recent experimental demonstrations include ultrafast phononic manipulation of magnetic orders~\cite{stremoukhov2022phononic, stupakiewicz2021ultrafast, nova2017effective}, of insulator to metal phase transitions~\cite{rini2007control, Caviglia2012Ultrafast, horstmann2020coherent}, and a transient enhancement of martensitic phase~\cite{Song2023UltrafastMP} and superconducting correlations~\cite{kaiser2014optically, hu2014optically, mankowskyNonlinearLatticeDynamics2014, mitrano2016possible}. In topological quantum materials, coherent phonon excitations were shown to induce dynamical switching between different topological phases by modifying the crystal symmetry and by tuning strain fields~\cite{garate2013phonon, Saha2014PhononTI, Kim2015TopologicalPT, Wang2017PhononPT, Weber2018UsingCP, sie2019ultrafast, Vaswani2019LightDrivenRC, luo2021light, Wang2021ExpansiveOF}. Particularly, recent coherent phonon pumping work provides compelling evidence of light-induced Dirac points~\cite{Vaswani2019LightDrivenRC}, Weyl nodes~\cite{luo2021light} and enhanced stability of topological systems~\cite{Yang2020lLightCS, Yang2018TerahertzQT}.

The theoretical understanding and first-principles simulations of light-excited electron-ion quantum systems are challenging, but significant progress has been achieved in recent years~\cite{prezhdo2021modeling, lindh2020quantum, nelson2020non, curchod2018ab, caruso2023quantum}. Specifically to describe the ultrafast electronic and spin dynamics associated with phonon excitations, microscopic theories have been developed for the light-induced insulator-to-metal structural phase transition~\cite{subedi2014theory}, the switching of magnetic orders~\cite{gu2018nonlinear}, and the enhancement of superconducting correlations  through symmetry-allowed electron-phonon coupling~\cite{rainesEnhancementSuperconductivityPeriodic2015b, komnikBCSTheoryDriven2016a, knapDynamicalCooperPairing2016, patelLightinducedEnhancementSuperconductivity2016, kennes2017transient, babadi2017theory, mazza2017nonequilibrium, murakamiNonequilibriumSteadyStates2017b, sentefLightenhancedElectronphononCoupling2017, Schuett-PRB-2018}. Here, we theoretically investigate the switching between strong and weak topological insulators (STI and WTI) induced by THz-driven coherent Raman phonon excitations in the model Dirac system ZrTe$_5$~\cite{Vaswani2019LightDrivenRC}. Although a qualitative picture has previously been established using static density functional theory (DFT) calculations~\cite{Vaswani2019LightDrivenRC}, the ultrafast dynamics of the laser-driven system such as the observed continuous increase of electronic carrier density after the THz pump and the underlying mechanism, calls for a more in-depth quantum dynamics simulations. The progress in this direction is highly appealing to the experimental community, who have been actively pursuing THz-driven quantum dynamics in various quantum materials recently~\cite{Luo2023QuantumCT, Kim2021TerahertzNI, Liu2020UltrafastCE, Liu2020CoherentBE}.

In this paper, we simulate the coherent phonon-induced carrier dynamics in ZrTe$_5$ in the framework of time-dependent Schr\"odinger equation with DFT basis functions. We complement the DFT-based dynamics simulations by an effective model calculations that captures the essentials of the microscopic mechanism. Our detailed numerical analysis shows that the switching between STI and WTI, which necessarily involves the closing of the bulk gap, creates a small but finite volume in momentum space, where effective two-level systems (TLSs) undergo avoided level crossings. This results in Landau-Zener-St\"uckelberg (LZS) tunnelling ~\cite{landauZurTheorieEnergieubertragung1932,zenerNonadiabaticCrossingEnergy1932,stuckelbergTheorieUnelastischenStosse1932,majoranaAtomiOrientatiCampo1932, shevchenko2010landau, ivakhnenko2023nonadiabatic} and leads to an increase of the carrier concentration during several cycles of the coherent phonon modulation. Our time-dependent Schr\"odinger equation simulations predict the dynamics of the phonon-induced carrier concentration in quantitative agreement with experiment. 

\section{Results and discussion}
\subsection{Summary of previous pump-probe experimental results}
\label{sec:pump_probe_experimental_results}
\begin{figure}[h]
	\centering
	\includegraphics[width=\columnwidth]{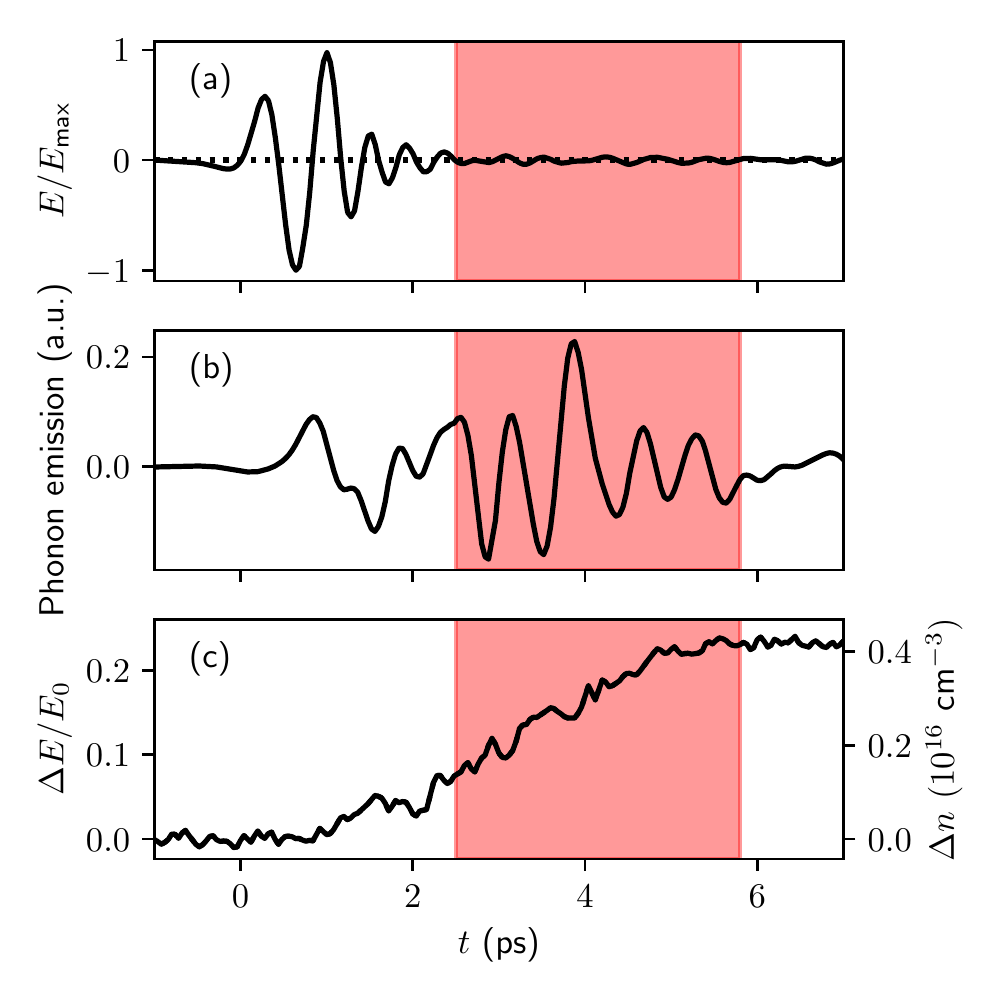}
	\caption{
	\textbf{THz pump-THz probe experimental results of ZrTe$_5$ at $4.2$ K.} (a) Normalized THz pump $E$-field as a function of pump delay time $t$, with the maximal value of $E(t)$ being $E_\text{max} = 736$~kV/cm. (b) Phonon emission as a function of $t$. The time-period after the pump pulse, where a coherent phonon emission is observed, is highlighted in red. This region is the focus of this work. (c) Normalized THz probe differential transmission $\Delta E (t)/E_0$ as a function of $t$. $\Delta E (t)$ is the differential transmission of the THz probe peak field $E_0$ measured with and without the THz pump pulse. The right $y$-axis labels the corresponding change of the carrier density $\Delta n$ in unit of $10^{16}$ cm$^{-3}$. 
	}
	\label{fig: expt}
\end{figure}

To facilitate the presentation, we summarize the key results of the THz pump-THz probe experiment on ZrTe$_5$ at $4.2$ K, which is described in detail in Ref.~\cite{Vaswani2019LightDrivenRC}. These results motivate the numerical simulations in this work. In the experiment, an intense THz-pump pulse with an $E$-field trace plotted in Fig.~\ref{fig: expt}\hyperref[fig: expt]{(a)} is incident normally on the ZrTe$_5$ single crystal. The THz pump-induced coherent phonon emission from the sample is observed after the pump pulse between $2.5$~ps~$ \lesssim t \lesssim 5.8$~ps, as highlighted in Fig.~\ref{fig: expt}\hyperref[fig: expt]{(b)}. The coherent phonon emission lasts for about five full cycles and its dominant spectral peak at $f_\text{ph}=1.2$~THz after Fourier transformation matches the $A_{1g}$ Raman mode at the Brillouin zone center. By performing THz pump and THz probe measurements using the same pump pulse, the THz probe differential transmission $\Delta E/E_0$, which is proportional to the change of carrier density $\Delta n$, is obtained and plotted in Fig.~\ref{fig: expt}\hyperref[fig: expt]{(c)}. The carrier density continuously increases after the pump pulse as long as the coherent phonon emission is observed. It saturates for $t \gtrsim 5.8$~ps, which coincides with the loss of phonon coherence. $\Delta E/E_0$ decays back to zero after about $120$~ps~\cite{Vaswani2019LightDrivenRC}.

In this paper, we focus on simulating this intriguing carrier excitation dynamics for the time period $2.5$~ps~$ \lesssim t \lesssim 5.8$~ps, where the coherent phonon excitation is present. The residual pump pulse is negligible during this time period, and the coherent phonon excitation can be treated as preexisting, i.e., without explicitly modelling the light-driven phonon generation process. The strong correlation between the carrier generation and the coherent phonon emission suggests a charge excitation mechanism assisted by a coherent Raman vibration. Indeed, by adiabatically following the $A_{1g}$ phonon trajectory, static DFT calculations have revealed that the electronic state of the system undergoes a topological transition between STI to critical Dirac point (DP) to WTI~\cite{Vaswani2019LightDrivenRC} (see also Appendix~\ref{sec: DFT}). This suggests the importance of the associated closing of the bulk band gap and potentially further topological effects in the carrier pumping process~\cite{Vaswani2019LightDrivenRC}. This makes a detailed quantum dynamics simulations of the physical process highly desirable.

\subsection{Effective model description}
\label{sec:tb_model}
\textbf{Model setup.}
\label{subsec:model_setup}
We first study a toy model that qualitatively captures the dynamical carrier generation observed experimentally in ZrTe$_5$. This model includes the essential physics underlying this phenomenon which is the topological phase transition (driven by coherent phonon oscillations) and a resulting inter-band charge excitation. For simplicity, we consider a Kitaev chain model~\cite{TopologyCMP}, which in momentum space is represented by the following two-orbital spinless fermion Bogoliubov-de Gennes (BdG) Hamiltonian:
\be
H(k) = (-2\nu\cos(k) - \mu)\tau_z + 2\Delta \sin(k) \tau_y\,. \label{eq: h_static}
\ee
Here $\tau_i$ are the Pauli matrices and the Hamiltonian parameters include onsite energy $\mu$, nearest-neighbor hopping $\nu$ and a superconducting pairing amplitude $\Delta$. The momentum lies in the range $k\in [-\pi, \pi)$. The model obeys particle hole symmetry $\tau_x H^*(-k) \tau_x = -H(k)$. 

\begin{figure*}[t!]
	\centering
 \includegraphics[width=\textwidth]{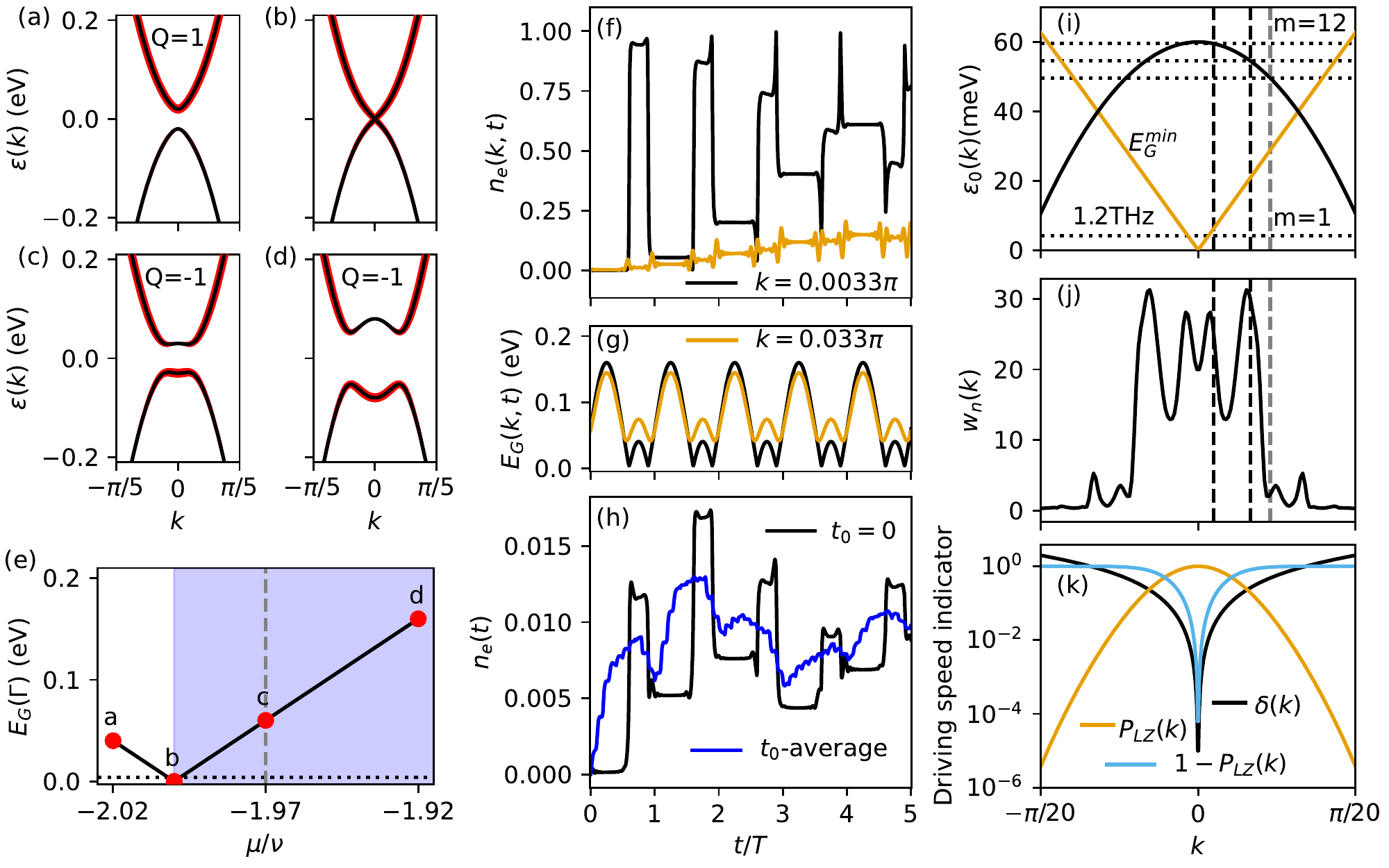}
	\caption{
	\textbf{Toy model results of phonon-induced topological phase transition and carrier excitation dynamics.} (a-d) Band structure of the BdG Kitaev chain model with $\mu/\nu = -2.02$, $-2.00$, $-1.97$, and $-1.92$. The red color encodes the projection weight of the band wavefunction on the first basis orbital ($\tau_z = 1$). The topological index $Q=\pm 1$ is also shown. (e) Band gap at the zone center $\Gamma$ point, $E_G(\Gamma)$, as a function of $\mu/\nu \in [-2.02, -1.92]$. This behavior qualitatively mirrors the behavior of ZrTe$_5$ system under the $A_\text{1g}$ Raman phonon modulation~\cite{Vaswani2019LightDrivenRC}. Red circles indicate $\mu/\nu$ values in panels (a-d). The topological region is highlighted in blue. The vertical dashed line indicates the equilibrium value $\mu_0$. (f) Time evolution of the excited state population $n_e (k, t)$ for a periodic modulation $\mu(t)/\nu \in [-2.02, -1.92]$ starting at $\mu(t=0) = \mu_0$ for five full cycles. The black line is for $k=0.0033\pi$ and the orange line for $k=0.033\pi$. (g) Time evolution of the energy gap $E_G(k, t)$ at $k=0.0033\pi$ (black) and $0.033\pi$ (orange). (h) Excited state population $n_e (t)$. The black curve is for $t_0 = 0$ and the blue one is averaged over $10$ runs with $t_0$ taken from $10$ uniformly spaced points in the interval $[0, T)$. (i) The $k$-point dependence  of the offset $\varepsilon_0(k)$~\eqref{eq: offset}. The bottom horizontal dotted line in (e, i) indicates the energy of the $f_\text{ph} = 1.2$~THz $A_\text{1g}$ phonon $E_\text{ph}$. The upper three horizontal dotted lines in (i) correspond to $E=m E_\text{ph}$ with $m=10$, $11$, and $12$. The vertical dashed lines in panels (i,j) indicate the $k$-points where $\varepsilon_0(k) = m E_\text{ph}$. The minimal energy gap $E_G^\text{min}(k)\equiv \min_{t \in [0,T]} E_G(k, t) = \abs{\Delta(k)}$ over the simulation period is also plotted as orange line for reference. (j) The $k$-point resolved number density in the excited band, which is defined as $W_n(k) \equiv \int_0^{5T} n_e(k, t) dt /\int_0^{5T} n_e(t) dt$. 
    %The momentum space region where a diabatic level crossing occurs is highlighted in blue, where the diabatic state is defined as $\ket{\psi(k, t=0)}$. 
    (k) The $k$-dependent driving speed indicators: $\delta(k)$ in black line, LZ transition probability $P_\text{LZ}(k)\equiv e^{-2\pi\delta(k)}$ in orange line, and $1-P_\text{LZ}(k)$ in sky blue line. 
	}
	\label{fig: tls}
\end{figure*}

With the chemical potential fixed at zero, the model exhibits a topological phase transition from a gapped superconductor that is trivial (NI) to one that is topological (TI) by tuning $\mu$. The corresponding BdG band structures together with the orbital ($\tau_z$) projections of the wavefunction are shown in Fig.~\ref{fig: tls}\hyperref[fig: tls]{(a-e)}. This evolution of the band energies is qualitatively similar to the phonon-induced topological phase transition in ZrTe$_5$ obtained from DFT calculations~\cite{Vaswani2019LightDrivenRC, Aryal2021TopologicalPT}. The band structure in panels (a) to (d) are obtained for $\mu/\nu$ equal to $-2.02$, $-2.00$, $-1.97$ to $-1.92$ (a-d). The evolution of the band gap as a function of $\mu/\nu$ is shown in panel (e). The bands in panels (a-d) are plotted together with weight of the projection of the Bloch wave function onto the first basis orbital $(\tau_z = 1)$, as given by the size of the red circles. Clearly, a band inversion occurs when the system transforms from the trivial phase ($\mu/\nu < -2$) to the topological phase ($\mu/\nu > -2$). At $\mu/\nu = -2$, the band gap closes and a nodal point forms at $\Gamma$ point. Consistently, the topological index, which can be defined as $Q \equiv \text{sign}(\mu^2-4\nu^2)$~\cite{TopologyCMP}, is $1$ in the trivial and $-1$ in the topological phase, as labelled in panels (a,c,d). For the numerical simulations, we set $\nu=1$~eV, and $\Delta/\nu = 0.1$, which results in a mode speed $\partial \varepsilon_k / \partial k = \pm 2\Delta$ at the Dirac point ($\mu/\nu=-2$) that is in qualitative agreement with the Fermi velocity in ZrTe$_5$. 

To model the effect of the coherent phonon excitation in ZrTe$_5$, we consider a periodic modulation of the onsite energy which resembles the treatment of electron-phonon coupling in the Holstein model~\cite{holstein1959studies}:
\begin{equation}
    \mu(t) = \mu_0 + \mu_1 \sin[\omega (t-t_0)]
\end{equation}
resulting in the time-dependent Hamiltonian
\be
H(k, t) = [-2\nu\cos(k) - \mu(t)]\tau_z + 2\Delta \sin(k) \tau_y\,. \label{eq: h_dyn}
\ee
We choose $\omega/2\pi=1.2$~THz ($T=833$~fs, $\hbar \omega = 4.96$~meV) to match the experimental value of the $A_\text{1g}$ Raman mode frequency. We set $\mu_0/\nu = -1.97$ and $\mu_1/\nu = 0.05$, such that the gap variation at the $\Gamma$ point is approximately the same as in ZrTe$_5$~\cite{Vaswani2019LightDrivenRC}, as shown in Fig.~\ref{fig: tls}\hyperref[fig: tls]{(e)}. Because the zone-center phonon carries zero momentum ($q=0$) it does not mix different Bloch momenta and the Hamiltonian $H(k, t)$ thus remains block diagonal in momentum space. The quantum dynamics simulation can therefore be performed by solving the time-dependent Schr\"odinger equation separately at each $k$-point:
\be
i\hbar \frac{\partial}{\partial t} \ket{\psi(k, t)} = H(k, t) \ket{\psi(k, t)}\,.
\label{eq:SE_toy_model}
\ee
We adopt a discrete-time propagator based on a Trotter decomposition of the state evolution~\cite{trotter, nielsen2002quantum} 
\be
\ket{\psi(k, t+ dt)} = e^{-i \frac{H(k, t)}{\hbar}dt} \ket{\psi(k, t)},
\ee
where the time step $dt \ll T$ is chosen sufficiently small compared to the variation of the onsite energy. 

\textbf{Dynamics simulation results.}
\label{sec: model_results}
The simulation starts at time $t=t_0$ and evolves until time $t=t_0 + 5T$ to agree with the experimental situation (see Fig.~\ref{fig: expt}). We monitor the time-dependent excited state population $n_e(t)$, which corresponds to the excited carrier density in the experiment and is tied to the differential emission $\Delta E/E$ that is measured experimentally. In our simulations we obtain $n_e(t) = \frac{1}{2\pi}\int_{-\pi}^\pi dk n_e(k,t) \approx \sum_k w_k n_e(k, t)$ as a weighted sum of contributions $n_e(k, t)$ at each $k$-point in the Brillouin zone $[-\pi, \pi)$. The weight $w_k$ is obtained as $1/N_k$ with $N_k$ the total number of $k$-points uniformly sampled in the Brillouin zone. Here we define 
\be
n_e(k, t) = \abs{\ov{\psi(k, t)}{\psi_c(k, t)}}^2, \label{eq: nekt}
\ee
which is the size of the projection of the one-electron wavefunction $\psi(k, t)$ on the adiabatic conduction band $\psi_c(k, t)$, an eigenstate of $H(k, t)$ with eigenvalue $\epsilon_c(k, t)$. We use a step size $dt=T/1000=0.833$~fs and a uniform $k$-mesh with $N_k = 560$ points, and find that $n_e(t)$ converges to a precision of $10^{-6}$. 

In Fig.~\ref{fig: tls}\hyperref[fig: tls]{(f)} we plot the excited state population $n_e(k, t)$ as a function of time $t$. We consider five modulation cycles of the onsite energy $\mu(t)$ starting with $t_0 = 0$. The black line represents the data at a momentum point close to the zone center $k=0.0033\pi$, and the orange line is for momentum $k=0.033\pi$. The corresponding (instantaneous) energy gap $E_G(k, t)$ at the respective $k$-points versus $t$ is shown in Fig.~\ref{fig: tls}\hyperref[fig: tls]{(g)}. Sharp variations of $n_e(k, t)$ are observed whenever the energy gap is minimal. In total we observe a substantial increase of $n_e(k, t)$ from zero to a finite value at the two representative $k$-points during the dynamical process. We note that the definition of $n_e(k, t)$~\eqref{eq: nekt} may not be unique. In Supplementary Note 1, we compare the numerical result with that based on an alternative definition, and show that the definition~\eqref{eq: nekt} gives more physically reasonable results.

The excited state population per unit cell, $n_e(t)$, is plotted as a black line in Fig.~\ref{fig: tls}\hyperref[fig: tls]{(h)}. It shows a similar behavior as $n_{e}(k,t)$ at the two individual $k$-points shown in Fig.~\ref{fig: tls}\hyperref[fig: tls]{(f)}. By the end of the simulation, $n_e(t=t_0 + 5 T)$ has increased from zero to about $0.01$. To account for the experimentally unknown initial phase of the coherent phonon oscillation, we also provide results that are averaged over $t_0$ that is uniformly sampled within the interval $[0, T]$. The averaged results $\Bar{n}_e (t) = \frac{1}{10}\sum_{i=0}^{9} n_e(t)|_{ t_0=\frac{i}{10} T}$ are shown in orange in Fig.~\ref{fig: tls}\hyperref[fig: tls]{(h)} and also increase from zero to about $0.01$ during the simulations.

\textbf{Discussion of toy model results.}
\label{sec: model_analysis}
The dynamics of the wavefunction at different $k$-points is completely independent from each other [see Eq.~\eqref{eq:SE_toy_model}]. 
The above calculation is thus composed of a collection of independent and periodically-driven two-level systems (TLSs), which resembles the well-known problem of LZS tunneling of a driven TLS in the presence of an avoided crossing~\cite{shevchenko2010landau, ivakhnenko2023nonadiabatic}. The behavior of the TLS is largely determined by the potential ramp speed (i.e. the oscillation frequency of the drive) and the minimal energy gap in the avoided crossing. 
% Although there exist some subtle differences in the time-dependence of the models, the dynamics of the ZrTe$_5$ toy model can be understood in analogy to the LZS problem.  
%the potential ramp speed (i.e. the oscillation frequency of the drive) and the minimal energy gap in the avoided crossing, which determine the behavior of the TLS during LZS tunneling, are also essential in our simulations of the ZrTe$_5$ toy model. 
While the ramp speed is set by the phonon frequency ($1.2$~THz), the minimal energy gap $E_G^\text{min}(k) \equiv \min_{t \in [0,T]} E_G(k, t)$ and the nature of the diabatic level crossing is strongly $k$-point dependent. In Fig.~\ref{fig: tls}~\hyperref[fig: tls]{(i)}, we show the minimal gap $E_G^\text{min}$ versus $k$ in the range $k \in [-\pi/20, \pi/20]$. The minimal gap changes almost linearly from to zero in the zone center ($k=0$) to about $60$~meV at $k=\pm \pi/20$. For reference, we also plot the $1.2$~THz phonon mode energy ($E_\text{ph} = 4.96$~meV) as a dotted horizontal line. These findings suggest that the contribution to the excited carrier density $n_e(t)$ arises from a small part in momentum space around the zone center. 

To obtain a more quantitative analysis, we define the following time-averaged and $k$-resolved excited state density:
\begin{equation}
    W_n(k) \equiv \frac{\int_0^{5T} n_e(k, t) dt}{\int_0^{5T} n_e(t) dt}\,.
\end{equation}
We plot $W_n(k)$ in Fig.~\ref{fig: tls}~\hyperref[fig: tls]{(j)} for a simulation with $t_0=0$, and find that it peaks in a $\Gamma$-centered narrow $k$-region. Interestingly, peaks occur in a wider range than naively expected by the condition that $E_G^\text{min}(k) \le E_\text{ph}$. This shows that higher order LZS resonances are important, where excitations occur across a minimal band gap that is a multiple of the driving frequency. Following the analysis of LZS tunneling, we  define diabatic states $\ket{\Tilde{\psi}(k)} \equiv \ket{\psi(k, t=0)}$. We find that the $k$-range, where a diabatic level crossing occurs, $\Av{\Tilde{\psi}_0(k)}{H(k, t')} = \Av{\Tilde{\psi}_1(k)}{H(k, t')}$, at some time $0 < t' < 5T$, matches well with the region of significant $W_n(k)$ as highlighted in blue. 
%This manifests the crucial role of phonon-induced topology switching for carrier excitation, namely, it creates a finite momentum space volume with effective TLSs containing avoided level crossing and sufficiently small gap, such that carriers can be excited through LZT-type mechanism.
We observe several distinct peaks of $W_n(k)$ in this region: the location of the peak closest to $\Gamma$-point ($k=0.004\pi$) and that of the peak furthest away ($k=0.033\pi$) are close to the $k$-points chosen for the presentation of $n_e(k, t)$ in (f, g). The peaks can be understood by considering resonance condition of the multi-cycle LZS problem as detailed in the following paragraph.
%The peak at $k=0.004\pi$ can be explained by the minimal gap being resonant with the coherent phonon frequency: $E_G^\text{min}(k) = E_\text{ph}$. The $\Gamma$ point, where the band gap closes exactly, is a location of local minimum (but the value of $W_n$ is still substantial). As illustrated by the vertical dashed lines, the largest peak at $k=5\times 0.0033\pi$ corresponds to $E_G^\text{min}(k) = 5E_\text{ph}$, while the far right peak at $k=0.033\pi$ corresponds to $E_G^\text{min}(k) = 10E_\text{ph}$.
%Following the analysis of LZS tunneling, we define diabatic states $\ket{\Tilde{\psi}(k)} \equiv \ket{\psi(k, t=0)}$ and find that the $k$-range, where a diabatic level crossing occurs ($\Av{\Tilde{\psi}_0(k)}{H(k, t')} = \Av{\Tilde{\psi}_1(k)}{H(k, t')}$) at some $t'>0$ during the simulation, matches well with the region of significant $W_n(k)$ as highlighted in blue. 

Although there is no closed form for the solution of the general LZS problem, some analytical understanding can corroborate the numerical results~\cite{shevchenko2010landau, ivakhnenko2023nonadiabatic}. To facilitate the discussion, we cast the time-dependent Hamiltonian~\eqref{eq: h_dyn} into the following form adopted in references~\cite{shevchenko2010landau, ivakhnenko2023nonadiabatic}:
\be
H(k, t) = -\frac{\Delta(k)}{2} \tau_x -\frac{\varepsilon(k, t)}{2}\tau_z\,,
\ee
where we apply a $-\pi/2$ rotation around $z$-axis, and define $\Delta(k) \equiv -4\Delta \sin(k)$, $\varepsilon(k, t) \equiv \varepsilon_0(k) + A\sin[\omega(t-t_0)]$, with the offset:
\be
\varepsilon_0(k) = 4\nu\cos(k) +2\mu_0\,, \label{eq: offset}
\ee
and the amplitude $A = 2\mu_1$. The level crossing of the diabatic states (eigenstates of $\tau_z$) occurs at $\varepsilon(k, t)=0$, which requires $-0.09\pi \lesssim k \lesssim 0.09\pi$ for the specific parameters of the model we set. Since the occupation probability of the upper adiabatic state is known to be negligibly small if no diabatic level crossing occurs~\cite{shevchenko2010landau, ivakhnenko2023nonadiabatic}, this is consistent with the narrow $k$-range with significant electron population transfer as shown in Fig.~\ref{fig: tls}~\hyperref[fig: tls]{(j)}. 
The minimal energy gap between the two adiabatic (i.e. instantaneous) levels reads $E_G^\text{min}(k) = \abs{\Delta(k)} = 4 \Delta \abs{\sin k}$. It is reached simultaneously with the diabatic state level crossing, and shows a linear behavior at small $k$ as plotted in Fig.~\ref{fig: tls}~\hyperref[fig: tls]{(i)}. As detailed in references~\cite{shevchenko2010landau, ivakhnenko2023nonadiabatic}, one can obtain an approximate analytic solution of the LZS problem, in the slow and the fast-passage limits. These limits are characterized by the dimensionless parameter $\delta(k) \equiv [\Delta(k)]^2/[4v(k)]$, which characterizes the ratio of the minimal gap $\abs{\Delta(k)}$ and the velocity $v(k) \equiv \frac{\partial \varepsilon(k, t)}{\partial t} \vert_{t_c: \varepsilon(k, t_c)=0} = A\omega \sqrt{1-[\varepsilon_0(k)/A]^2}$ at the minimal gap. The velocity is finite and only weakly varying in the region of interest close the $k=0$ before it rapidly drops to zero as $k\rightarrow 0.09\pi$. %It experiences a local minimum at $k=0$ and a maximum at $k=0.06 \pi$ that is only 1.25 times larger. It then rapidly drops to zero as $k\rightarrow 0.09\pi$ at the edge of the region with the diabatic level crossing. 
As a result, we find $\delta(k) \propto k^2$ at small $k$ and $\delta = 1$ at $k_1 \approx 0.03 \pi$ (see Fig.~\ref{fig: tls}~\hyperref[fig: tls]{(k)}). In terms of this parameter, the LZ transition probability for a single passage reads $P_\text{LZ}(k)\equiv e^{-2\pi\delta(k)}$. In the fast-passage limit at $k \ll k_1$, we have $\delta(k) \ll 1$ and $1-P_\text{LZ}(k) \ll 1$; while in the slow-passage limit, we have $\delta(k) \gg 1$ and $P_\text{LZ}(k) \ll 1$. 
In Fig.~\ref{fig: tls}~\hyperref[fig: tls]{(k)} we plot these driving speed indicators in the $k$-region with significant excited state occupation. The fast-passage regime extends from the $\Gamma$-point ($k=0$), where the minimal gap vanishes, roughly to the position of the first peak of $W_n(k)$ at $k=0.0033\pi$ [see Fig.~\ref{fig: tls}~\hyperref[fig: tls]{(k)}], where $\delta(k) \lesssim 0.01$ and $1-P_\text{LZ}(k) \lesssim 0.06$. The resonance condition in the diabatic regime reads $\varepsilon_0(k) = m \omega$ with integer $m$. This condition is fulfilled at $k=0.005 \pi$ for $m=12$ and at $k=0.017\pi$ for $m=11$ as shown in by the black vertical dashed lines in Fig.~\ref{fig: tls}~\hyperref[fig: tls]{(i)}, which explains the two dominant peaks seen in Fig.~\ref{fig: tls}~\hyperref[fig: tls]{(j)}. Note that we have $\delta(k=0.017\pi) = 0.26$ and $1-P_\text{LZ}(k=0.017\pi) = 0.81$, which implies that $k=0.017\pi$ is near the crossover from fast to slow passage. In fact, the resonance condition with $m=10$ gives $k=0.023\pi$ as shown by a gray vertical dashed line in Fig.~\ref{fig: tls}~\hyperref[fig: tls]{(i)}, which is off from the peaks in Fig.~\ref{fig: tls}~\hyperref[fig: tls]{(j)} due to being located close to the crossover region with $\delta(k=0.023\pi) = 0.48$ and $1-P_\text{LZ}(k=0.023\pi) = 0.95$. At larger $k_1 \approx 0.03 \pi$, the crossover from fast to slow passage occurs and one needs to use a more general resonance condition (see Eq.~(56) in Ref.~\cite{ivakhnenko2023nonadiabatic}). Finally, for $\pi/20 \leq k \leq 0.09 \pi$ and beyond the slow-passage regime is reached, where we find a small adiabatic state probability. The resonance condition in the adiabatic regime reads $\frac{2 A}{\pi \omega} = m$ with integer $m$, which is not exactly fulfilled for our choice of parameters $\frac{2 A}{\pi \omega} = 12.8$. One should also take into account that the dynamics only involves five LZS oscillations due to damping effects in experiments, which limits the total transfer into the upper adiabatic state (even on resonance) when $P_{\text{LZ}} \ll 1$.
%around $k=k_1=0.03 \pi$ and near $k=\pi/20$ shown in Fig.~\ref{fig: tls}~\hyperref[fig: tls]{(k)} and beyond, the slow-passage limit is reached.
%, as $\delta(k) \gtrsim 2$ and $P_\text{LZ}(k) \lesssim 10^{-6}$. 
We conclude that the majority of the excited state population dynamics in our model occurs in the fast-passage region and the crossover regime between the fast and the slow-passage limits, and that the main peaks can be understood as arising from resonances in the diabatic regime. 
% There is no closed form of the solution in the intermediate regime. 
% Note that in fast-passage limit, the resonance condition is given by $\Tilde{\mu}_0(k)/\omega = m$, where $m$ is an integer. We find that $\Tilde{\mu}_0(k)/\omega \approx 14.6$ in our study, which indicates that the dynamics of the model is not in resonance condition in the narrow zone-center region where the fast-passage limit holds. 

To summarize, our analysis demonstrates the crucial role of the phonon-induced topological band closing for carrier excitation. 
%The band closing is associated with the change of the ground state topology. , switching for carrier excitation, namely, 
This creates a finite momentum space volume where effective TLSs experience an avoided level crossing with a sufficiently small band gap such that carriers can be excited through LZS tunneling.

\subsection{First-principles quantum dynamics simulations}
\label{sec:first_principles_simulations}
\textbf{Model and ab initio simulation method.}
\label{subsec:methods}
To gain a more material-specific understanding of the carrier excitation dynamics of the phonon-modulated ZrTe$_5$ system, we carry out first-principles simulations based on time-dependent Schr\"odinger equation with DFT basis functions. The time-dependence of the KS Hamiltonian $\h_0 (t)$ is encoded in the ionic trajectory $\bR(t)$ that is set by the $A_{1g}$ coherent phonon. In the implementation of DFT for periodic systems one often adopts a basis set with large dimension, such as plane waves. This renders a direct manipulation of the DFT Hamiltonian $\h_0 (t)$ cumbersome. Note that $\h_0 (t)$ generally covers higher-energy unoccupied states and deeper occupied states, which are likely irrelevant for the carrier excitation dynamics in phonon-modulated ZrTe$_5$, which we expect to be dominated by states close to the chemical potential. Standard tight-binding downfolding approaches, such as maximally localized Wannier function~\cite{marzari2012maximally} and quasi-atomic minimal basis-set orbitals method~\cite{chan2007highly, qian2008quasiatomic}, can be useful; but the downfolding calculation for many snapshots along the trajectory $\bR(t)$ in the simulation time period can be time-consuming, and the time-dependence of the downfolded orbitals introduces additional complexity. Here we adopt an alternative representation where the component of the dynamical electronic state $\ket{\Psi(t)}=\otimes_{\bk}\ket{\Psi(\bk,t)}$ is approximated as a linear combination of $N_\text{b}$ adiabatic states $\{\ket{\Phi_i (\bk, t)}\}$ generalized to a generic $\bk$-point from Ref.~\cite{granucci2012surface, li2018spin}:
\be
\ket{\Psi(\bk, t)} = \sum_{i=1}^{N_\text{b}} c_i(\bk, t)\ket{\Phi_i(\bk, t)}, \label{eq: wf}
\ee
where $\ket{\Phi_i(\bk, t)} \equiv \prod_{\mu\in S_i}\phi_{\mu}^\dag (\bk, t) \ket{0}$ is a noninteracting single Slater determinant state defined by a set $S_i$ of occupied KS orbitals $\phi_{\mu}(\bk, t)$, which satisfies $\h_0(\bk, \bR(t))\ket{\phi_{\mu}(\bk, t)} = \epsilon_{\mu}(\bk, t)\ket{\phi_{\mu}(\bk, t)}$. Here the crystal momentum $\bk$ is conjugate to the position vector of the simulation unit cell.

The propagation of $\ket{\Psi(\bk, t)}$ is encoded in the time-dependent complex amplitudes $c_i(\bk, t)$ and the adiabatic states $\ket{\Phi_i(\bk, t)}$. Substituting Eq.~\eqref{eq: wf} into the time-dependent Schr\"odinger equation leads to the equation of motion (EOM) of the amplitudes
\be
i\hbar \frac{\partial c_i(\bk, t)}{\partial t} = \sum_{j=1}^{N_b}H_{i j}(\bk, t)c_j(\bk, t)\,. \label{eq: eom}
\ee
The vibronic Hamiltonian is given by
\be
H_{i j}(\bk, t) = \varepsilon_i(\bk, t) \delta_{i j} -i\hbar d_{i j}(\bk, t)\,. \label{eq: vh}
\ee
Here, we define $\varepsilon_i(\bk, t) = \sum_{\mu\in S_i}\epsilon_{\mu}(\bk, t) $. The complex nonadiabatic coupling (NAC) coefficient between a pair of distinct states $\{\ket{\Phi_i}, \ket{\Phi_j} \}$ is given by $d_{i j}=\Mel{\Phi_i}{\frac{\partial}{ \partial t}}{\Phi_j}$, which is nonzero only if there is exactly one distinct occupied KS orbital between $\ket{\Phi_i}$ and $\ket{\Phi_j}$ due to the single Slater determinant nature~\cite{HammesSchiffer1994ProtonTI, akimov2013pyxaid}. The NAC can be conveniently evaluated using the finite-difference method~\cite{HammesSchiffer1994ProtonTI}:
\bea
d_{i j}(\bk, t) \approx \frac{1}{2dt} &&\left( \ov{\Phi_i(\bk, t)}{\Phi_j(\bk, t+dt)}\right. \nonumber \\
&&\left. - \ov{\Phi_i(\bk, t+dt)}{\Phi_j(\bk, t)} \right), ~\label{eq: dij}
\eea
which is completely determined by the state overlap matrix between consecutive time steps.

\begin{figure*}[t!]
	\centering
	\includegraphics[width=\textwidth]{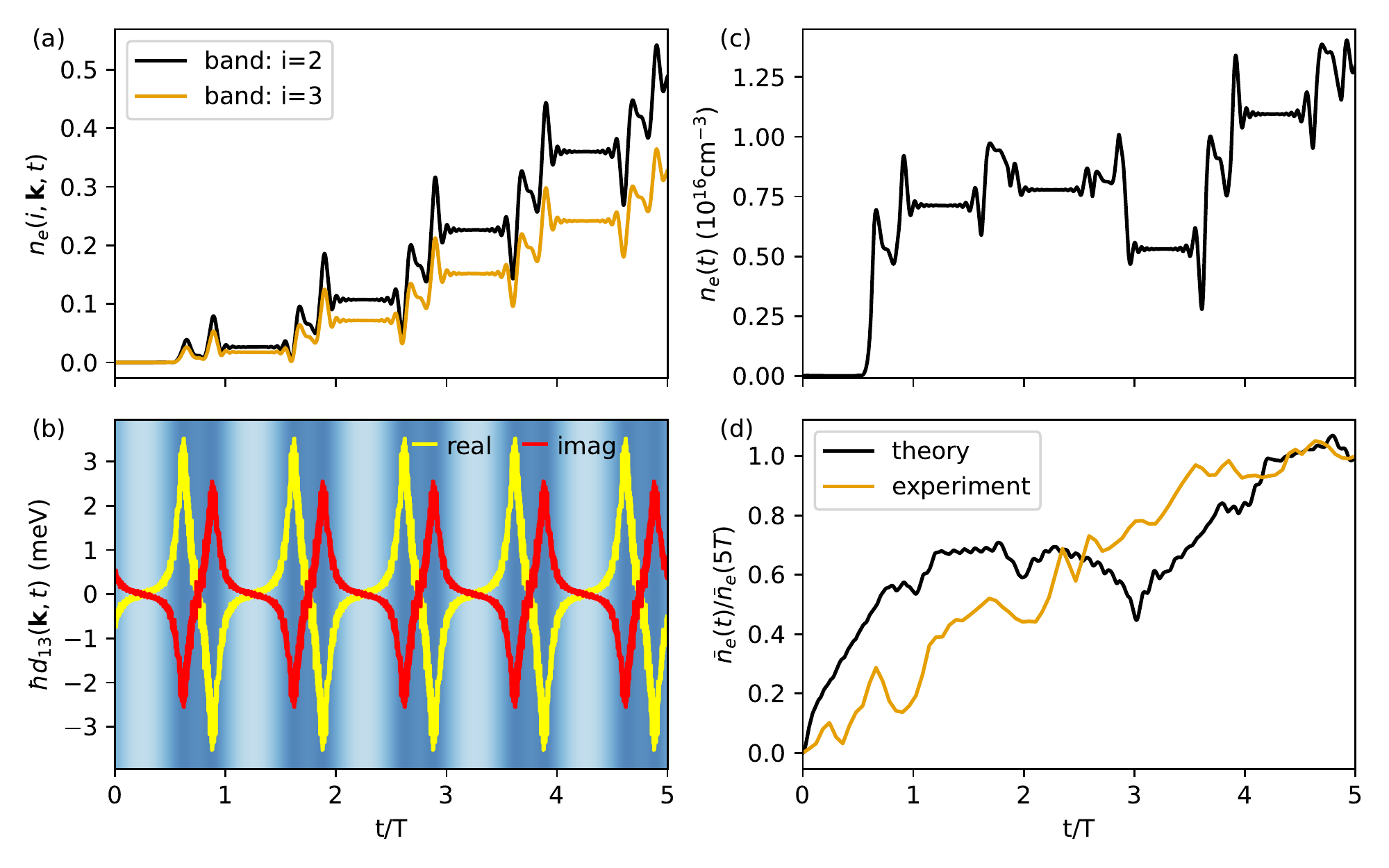}
	\caption{
	\textbf{Carrier excitation dynamics of phonon-modulated ZrTe$_5$ system from first-principles simulations.} (a) Occupancy of the two lowest-energy conduction bands, $n_e(i, \bk, t)$ with $i=2,3$, as a function of simulation time $t$ for $5$ phonon periods starting with $t_0=0$ ( the equilibrium configuration with zero phonon displacement) at $\bk=\mathbf{b}_1/720 + \mathbf{b}_2/720 + \mathbf{b}_3/80$. Here $(\mathbf{b}_1, \mathbf{b}_2, \mathbf{b}_3)$ are the reciprocal primitive vectors (see Appendix~\ref{sec: DFT} for details). (b) Time-dependence of the complex NAC amplitude $\hbar d_{13} (\bk, t)$ between the top valence band ($i=1$) and lowest conduction band ($i=3$) at the same $\bk$-point as (a), with real part in yellow and imaginary part in red. The background dark blue to light blue indicates a band gap, $E_\text{G} = \varepsilon_2 (\bk, t) - \varepsilon_1 (\bk, t)$, varying from $27$ meV to $132$ meV. (c) The total excited states population $n_e(t)$ integrated over the Brillouin zone as a function of $t$ for the simulation starting with $t_0=0$. (d) The time-evolution of normalized $\Bar{n}_e(t)/\Bar{n}_e(5T)$(black line), where $\Bar{n}_e(t)$ is the total excited states population $n_e(t)$ averaged over simulations starting with $t_0$ uniformly sampled by $10$-points in a phonon period. The normalized experimental carrier density change $\Delta n(t)/\Delta n(5T)$ are also plotted in orange line. We obtain $\Bar{n}_e(5T)=1.2\times 10^{16}$cm$^{-3}$ from the theoretical simulation, and $\Delta n(5T)=(0.28\pm 0.12)\times 10^{-16}$ cm$^{-3}$ estimated from the experiment.
	}
	\label{fig: zrte5}
\end{figure*}

\textbf{Technical details of the simulation.}
\label{subsec:technical_details}
The computational complexity of the simulation is tied to the number $N_\text{b}$ of adiabatic states that are used in Eq.~\eqref{eq: wf}. The Raman $A_{1g}$ phonon in ZrTe$_5$ has a frequency of $1.2$ THz, which is one order of magnitude smaller than the equilibrium band gap. We can thus truncate the expansion at the level of single-electron excitations between low energy bands. We include configurations with one electron excited from the top valence band to one of the four lowest conduction bands. We thus keep $N_\text{b}=5$ adiabatic states $\Phi_i(\bk, t)$ in the expansion at each $\bk$-point. Each $\Phi_i$ differs by exactly one occupied KS orbital, which therefore can also be labelled by the same index $i$, with $i=1$ corresponding to the top valence band, and $i=2\dots N_\text{b}$ to the conduction bands in ascending order of energy. The vibronic Hamiltonian~\eqref{eq: vh} can be simplified by setting $\varepsilon_i(\bk, t)=\epsilon_i(\bk, t)$ as a rigid potential shift. The evaluation of the NAC~\eqref{eq: dij} can also be reduced to
\bea
d_{i j}(\bk, t) \approx \frac{1}{2dt} &&\left( \ov{\phi_i(\bk, t)}{\phi_j(\bk, t+dt)}\right. \nonumber \\
&&\left. - \ov{\phi_i(\bk, t+dt)}{\phi_j(\bk, t)} \right)\,, 
\label{eq: dij_simp}
\eea
which depends only on the $N_\text{b}$ KS orbitals that are kept at each $\bk$-point.

The EOM~\eqref{eq: eom} assumes the continuity of the time dependent basis $\ket{\Phi_i(\bk, t)}$ with $t$. Therefore, it is crucial to fix the phase degree of freedom of the KS orbitals $\phi_i(\bk, t)$. This can be achieved by consecutively applying a phase factor to $\phi_i(\bk, t+dt)\to e^{i\theta}\phi_i(\bk, t+dt)$, where the phase $e^{i\theta} \equiv O_i^*/\abs{O_i}$ with $O_i = \ov{\phi_i(\bk, t)}{\phi_i(\bk, t+dt)}$ is determined by the overlap with the same orbital at the previous time step. Additional complexity of the dynamics simulations originates from the presence of time-reversal and inversion symmetry, which renders every band doubly degenerate, and strong spin-orbit coupling in ZrTe$_5$. The ambiguity in the doubly degenerate bands can be partially fixed by choosing a $S_z$-gauge such that the $2\times 2$ spin $S_z$ matrix becomes diagonal in each doubly degenerate manifold via a unitary transformation. Numerically, we find that the $S_z$ gauge transformation is not sufficient to guarantee the orbital continuity along the dynamical path. Therefore, we propose the following overlap gauge correction to better address the band degeneracy problem. The simulation starts with orbitals in the $S_z$-gauge, and apply unitary transformation in each doubly degenerate manifold in all following time steps, such that each rotated orbital $\phi_i (\bk, t+dt)$ has maximal overlap with the same one at the previous step. This is achieved by diagonalizing a series of $2\times2$ matrices $P_{r s}^{(j)} = \Mel{\phi_r^{(j)}(\bk, t+dt)}{\hat{P_j}}{\phi_s^{(j)}(\bk, t+dt)}$ with $\hat{P_j} = \ket{\phi_1^{(j)}(\bk, t)}\bra{\phi_1^{(j)}(\bk, t)}$. Here $r,s \in [1,2]$ run through the two orbitals in the $j$th degenerate doublet. The diagonalization gives two eigenvectors, where one has a nonzero eigenvalue and is assigned to the first orbital of the $j$th doublet. The other eigenvector has zero eigenvalue, and is assigned to the second orbital in the doublet. Following this procedure, we numerically find that the self-overlap of each wavefunction at consecutive time steps always remains above $99.99\%$. The band index exchange between different degenerate doublets, which can be detected by checking the overlap between wavefunctions at consecutive time steps, is not observed in the simulations reported here.

\textbf{First-principles quantum dynamics simulation results.}
\label{subsec:first_principles_results}
The analysis of the one-dimensional (1D) toy model results shows that the main contribution to the carriers in the excited band resulted from LZS tunneling in a narrow region of momentum space around the zone center $k \in [-\pi/20, \pi/20]$. Expecting a similar behavior for the realistic 3D model of ZrTe$_5$, we use a dense $360\times360\times120$ uniform $\bk$-grid covering the full Brillouin zone for the following dynamics simulations and use a shift of $\Delta = \frac{0.5}{360}(\mathbf{b}_1+\mathbf{b}_2) + \frac{0.5}{120}\mathbf{b}_3$ from the $\Gamma$-point for each $\bk$-point. Here, $\mathbf{b}_i$ are the reciprocal basis vectors. The center-shifted $\bk$-mesh therefore excludes the $\Gamma$-point, where the band gap closes at certain times and additional gauge correction is otherwise needed.

In Fig.~\ref{fig: zrte5}\hyperref[fig: zrte5]{(a)} we present the excited state population $n_e(i, \bk, t)$ as a function of simulation time $t$ for five phonon cycles, starting at $t_0=0$ with zero phonon displacement. Here we define 
\begin{equation}
    n_e(i, \bk, t) \equiv \abs{c_i(\bk, t)}^2\,,
\end{equation}
which is equivalent to the definition used in the toy model analysis. At momentum $\bk=\mathbf{b}_3/120 + \Delta$ adjacent to the zone center, an electron is gradually excited from the top valence band ($i=1$) to the two lowest conduction bands ($i=2, 3$) due to nonadiabatic effects. The complex NAC amplitude $\hbar d_{13} (\bk, t)$ between the $1$st and $3$rd bands at the same $\bk$-point is plotted in Fig.~\ref{fig: zrte5}\hyperref[fig: zrte5]{(b)}. The yellow line denotes the real part and the red line the imaginary part. The line width indicates the numerical noise, which is found to have negligible impact on the state population dynamics. The maximal difference in the dynamical state populations from the simulation using the (noisy) NAC amplitudes versus using smoothed data via application of a Savitzky–Golay filter is only about $10^{-3}$. Clearly, Fig.~\ref{fig: zrte5}\hyperref[fig: zrte5]{(a,b)} shows that a sharp transition of the state population $n_e$ occurs at the peaks of the NAC. This also coincides with a minimum of the band gap, $E_\text{G} = \varepsilon_2 (\bk, t) - \varepsilon_1 (\bk, t)$, as indicated by the blue shading in the background of Fig.~\ref{fig: zrte5}\hyperref[fig: zrte5]{(b)}. We note that $d_{12} (\bk, t)$ has a similar time-dependence as $d_{13} (\bk, t)$ [both bands have degenerate energies $\varepsilon_2 (\bk, t) = \varepsilon_3 (\bk, t)]$, yet with slightly larger amplitude. This difference in the NAC induces a larger electron occupancy in the $i=2$ band compared to the one with $i=3$, as shown in Fig.~\ref{fig: zrte5}\hyperref[fig: zrte5]{(a)}. In contrast, the NACs from $i=1$ to $i=4, 5$ are smaller by more than one order of magnitude, resulting in negligibly small carrier excitations to these bands, $n_e(i, \bk, t) < 10^{-6}$ for $i=4,5$. 

The first-principles dynamics simulations allow for a direct and quantitative comparison to experiment. First, in Fig.~\ref{fig: zrte5}\hyperref[fig: zrte5]{(c)} we show the excited state carrier density, $n_e(t) = \sum_{\bk} w_{\bk} \sum_{i=2}^{3} n_e(i, \bk, t)$ as a function of time $t$. It increases from zero to about $n_e(t=5T) \approx 2.4\times 10^{16}$cm$^{-3}$ at the end of the simulation $t = 5 T$ and exhibits qualitatively similar sharp transitions near dynamical band gap minimum as the momentum resolved quantity in Fig.~\ref{fig: zrte5}\hyperref[fig: zrte5]{(a)}. Within the $360\times360\times120$ uniform $\bk$-mesh of the Brillouin zone, we find that the dominant contributions come from $\bk = \frac{l}{360}\mathbf{b}_1 +\frac{m}{360}\mathbf{b}_2 + \frac{n}{120}\mathbf{b}_3 + \Delta$ with $l,m,n = 0, \pm 1, \pm 2, -3$. Next, we we account for the fact that in pump-probe experiments, the time-trace of differential transmission is obtained as an average over multiple runs and ZrTe$_5$ samples exhibit some degree of electronic heterogeneity and nanostrip junctions, as observed in the THz nanoimaging~\cite{kim2021terahertz}. To capture these phenomena on average, we define $\Bar{n}_e (t) = \frac{1}{10}\sum_{i=0}^{9} n_e(t)|_{ t_0=\frac{i}{10} T}$, which is an average over simulations at $10$ different starting times. In Fig.~\ref{fig: zrte5}~\hyperref[fig: zrte5]{(d)} we directly compare the time-dependence of a normalized $\Bar{n}_e (t)/\Bar{n}_e (t=5T)$ (black), to the experimental data $\Delta n (t)/\Delta n (t=5T)$ (orange). Both curves exhibit a similar growth pattern of the carrier density over time, and we also find the carrier density at the end of the simulation $\Bar{n}_e (t=5T) \approx 2.4\times 10^{16}$ cm$^{-3}$ to be in good agreement with the one estimated from experiment $\Delta n(5T) = (0.28\pm 0.12)\times 10^{-16}$ cm$^{-3}$, considering that there is electronic heterogeneity present in the experimental ZrTe$_5$ sample~\cite{kim2021terahertz}. The numerical estimation of $n_e(t=5T)$ and $\Bar{n}_e (t=5T)$ reported here also includes a factor of $2$ to take into account the double degeneracy of the top valence band in ZrTe$_5$ system.

\section{Conclusions}
\label{sec:conclusion}
We report detailed first-principle and effective model simulations of the carrier excitation dynamics in coherent phonon-modulated ZrTe$_5$. Our results shed new light on recent pump-probe experiments~\cite{Vaswani2019LightDrivenRC} by providing a clear intuitive explanation of the experimental results. Both first-principle and effective model calculations reveal the importance of the phonon-induced topological phase transition in ZrTe$_5$ and the associated closing of the bulk gap for the observed excitation of carriers. We show that the excitations occur via Landau-Zener-St\"uckelberg tunneling in a series of time-dependent avoided level crossings of Bloch states located in a narrow region of momentum space around the zone center. Our detailed time-dependent Schr\"odinger equation simulations further show that the dominant tunneling occurs between the highest valence band and the lowest doubly degenerate conduction bands, while excitations to the next higher bands are negligible due to small transition matrix elements. We predict that the carrier density increases gradually with time and reaches a final value of $2.4\times 10^{16}$cm$^{-3}$ at $t=5T$ when phonon coherence is lost in the experiment. These results are in good quantitative agreement with experiment. Our work thus demonstrates that the coherent charge excitation process in topological quantum materials such as ZrTe$_5$ can be understood and predicted quantitatively by first-principles quantum dynamics simulations.

\begin{figure*}[t!]
	\centering
	\includegraphics[width=\textwidth]{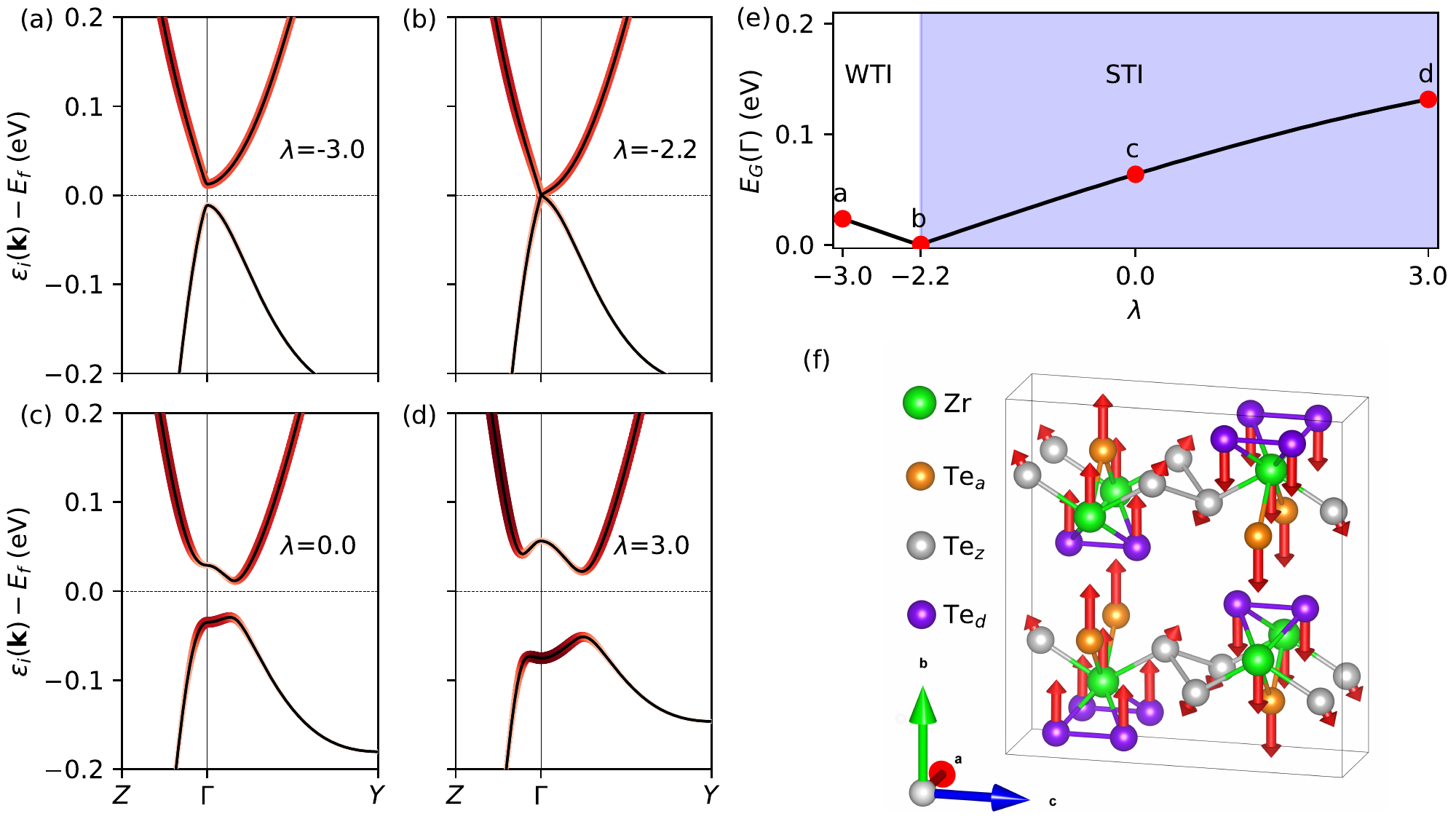}
	\caption{
	\textbf{DFT band structure calculations for the adiabatic topological phase switching induced by the $A_\text{1g}$ Raman phonon mode in ZrTe$_5$.} (a-d) Band structure along $\bk$-path $Z-\Gamma-Y$ with distortion parameter $\lambda = -3.0,-2.2, 0.0, and 3.0$, respectively. The red color encodes projection weight of $5p$-orbitals of Te$_d$ to the wavefunctions. We define $Z=(0, 0, 0.5)$ and $Y=(-0.5, 0.5, 0)$ in the reciprocal space with basis vectors $(\mathbf{b}_1, \mathbf{b}_2, \mathbf{b}_3)$. (e) Band gap at the zone center, $E_\text{G}(\Gamma)$, as a function of distortion parameter $\lambda$. Here $\lambda=1$ represented $0.033$~\AA displacement for Zr, $0.035$~\AA for Te$_d$, $0.032$~\AA for Te$_a$, and $0.017$~\AA for Te$_z$~\cite{Vaswani2019LightDrivenRC}. Red circles indicate the $\lambda$ values for panels (a-d). The STI region is highlighted in blue and WTI region in white, with boundary line indicating the Dirac point position. (f) The $A_\text{1g}$ phonon mode in the conventional cell of ZrTe$_5$. Green spheres represent Zr atoms, orange spheres for apical Te atoms (Te$_a$), silver spheres for zigzag Te atoms (Te$_z$), and purple spheres for dimerized Te atoms (Te$_d$). The arrows indicate the atomic displacement vectors of the $A_\text{1g}$ mode.
	}
	\label{fig: appendix}
\end{figure*}

\section{Methods}
\label{sec: DFT}
First-principles total energy and electronic-structure calculations for ZrTe$_5$ are based on DFT with the exchange correlation functional in generalized gradient approximation parametrized by Perdew, Burke, and Ernzerhof (PBE)~\cite{PBE}. Van der Waals interaction is included by Grimme's damped atom-pairwise dispersion corrections (D2)~\cite{grimme2006semiempirical}. The calculations are performed using the Vienna \textit{Ab initio} Simulation Package (VASP)~\cite{VASP}. We use a plane-wave cutoff energy of $230$~eV and include spin-orbit coupling for all the calculations. The phonon modes are calculated using the finite displacement approach as implemented in Phonopy~\cite{phonopy}. Specifically, we use the primitive unit cell of experimental structure($a=3.987$~\AA, $b=14.502$~\AA, and $c=13.727$~\AA)~\cite{matkovic1992constitutional}. The primitive vectors are $\mathbf{a}_1=(1.994, -7.251, 0)$~\AA, $\mathbf{a}_2=(1.994, 7.251, 0)$~\AA, and $\mathbf{a}_3=(0, 0, 13.727)$~\AA. The corresponding reciprocal primitive vectors are $\mathbf{b}_1=(0.251, -0.069, 0)2\pi$~\AA$^{-1}$, $\mathbf{b}_2=(0.251, 0.069, 0)2\pi$~\AA$^{-1}$, and $\mathbf{b}_3=(0, 0, 0.073)2\pi$~\AA$^{-1}$. Highly accurate wavefunctions at specific $\bk$-points are generated for NAC calculations by setting an energy convergence criterion to $10^{-9}$~eV and requiring a minimum of $20$ electronic steps.

To be self-contained, we present the key DFT band structure calculation results for the description of the $A_\text{1g}$ phonon-induced adiabatic topological phase transition observed in ZrTe$_5$ system as reported in Ref.~\cite{Vaswani2019LightDrivenRC}. Under the modulation of the $A_\text{1g}$ eigenmode as plotted in Fig.~\ref{fig: appendix}\hyperref[fig: appendix]{(f)}, the band gap $E_\text{G}(\Gamma)$ at zone center closes at distortion parameter $\lambda=-2.2$ as shown in Fig.~\ref{fig: appendix}\hyperref[fig: appendix]{(e)}, implying a topological phase transition along the dynamical path. This is confirmed by the band structure analysis and topological invariant index calculation~\cite{Vaswani2019LightDrivenRC}. In Fig.~\ref{fig: appendix}\hyperref[fig: appendix]{(a-d)}, we plot the band structure along high-symmetry $\bk$-path $Z-\Gamma-Y$ at phonon distortion parameter $\lambda=-3.0, -2.2, 0, 3.0$, decorated with red color indicating the Te$_d$ $5p$-orbital weight. Band inversion clearly occurs when $\lambda$ passes through $\lambda=-2.2$ the Dirac point. For $\lambda < -2.2$, the adiabatic state of the system is in WTI, and switches to STI for $\lambda > -2.2$. 

\section*{Data availability}
All the data to generate the ﬁgures are available at figshare~\cite{data_qd_zrte5}. Data supporting the calculations are available together with the codes at figshare~\cite{CyQuanDyn}. All other data are available from the corresponding authors on reasonable request.

\section*{Code availability}
All the computer codes developed and used in this work are available open-source at figshare~\cite{CyQuanDyn}.

\bibliography{ref}

\section*{Acknowledgements}
This work was supported by the U.S. Department of Energy (DOE), Office of Science, Basic Energy Sciences, Materials Science and Engineering Division, including the grant of computer time at the National Energy Research Scientific Computing Center (NERSC) in Berkeley, California. The research was performed at the Ames National Laboratory, which is operated for the U.S. DOE by Iowa State University under Contract No. DE-AC02-07CH11358. 

\section*{Author contributions}
T.J. performed the DFT and quantum dynamics simulations. Y.X.Y. and T.J. wrote the codes for dynamics simulation and analysis. P.P.O. initiated the effective model simulation and contributed to the analysis of the results. L.L. and J.W. performed the experimental analysis. L.L.W. and F.Z. helped with the DFT calculations and analysis. J.Z., C.Z.W., and K.M.H. provided inputs for the first-principle dynamics simulations. Y.X.Y., T.J., and P.P.O. wrote the paper, with contributions from all the authors. Y.X.Y. supervised the project.

\section*{Competing interests}
The authors declare no competing interests.

\let\oldtheequation\theequation
\let\theequation\thesuppeqn

\clearpage

%%%%%%%%%%%%%%%%%

%\widetext
\onecolumngrid
\begin{center}
\textbf{\large Supplemental materials for: \\ \textit{Ab-initio} Simulations of Coherent Phonon-Induced Pumping of Carriers in Zirconium Pentatelluride}
\vspace{0.4cm}
\end{center}
\twocolumngrid

%\noindent
%\begin{minipage}{\textwidth}
%  \begin{center}
%    \textbf{\large Supplemental material for: \\ Error mitigation in variational quantum eigensolvers using %tailored probabilistic machine learning}
%  \end{center}
%\end{minipage}

%%%%%%%%%% Prefix a "S" to all equations, figures, tables and reset the counter %%%%%%%%%%
\renewcommand{\figurename}{Fig.}
\renewcommand{\thefigure}{S\arabic{figure}}
\setcounter{equation}{0}
\setcounter{figure}{0}
\setcounter{section}{0}
\setcounter{table}{0}
\setcounter{page}{1}
\makeatletter

%remove these to get rid of S
%\renewcommand{\theequation}{S\arabic{equation}}
%\renewcommand{\thefigure}{S\arabic{figure}}
%\renewcommand{\bibnumfmt}[1]{[S#1]}
%\renewcommand{\citenumfont}[1]{S#1}\
%\renewcommand{\thesection}{S\arabic{section}}

\section{Alternative way for excited state population analysis}

\begin{figure*}
	\centering
 \includegraphics[width=0.6\linewidth]{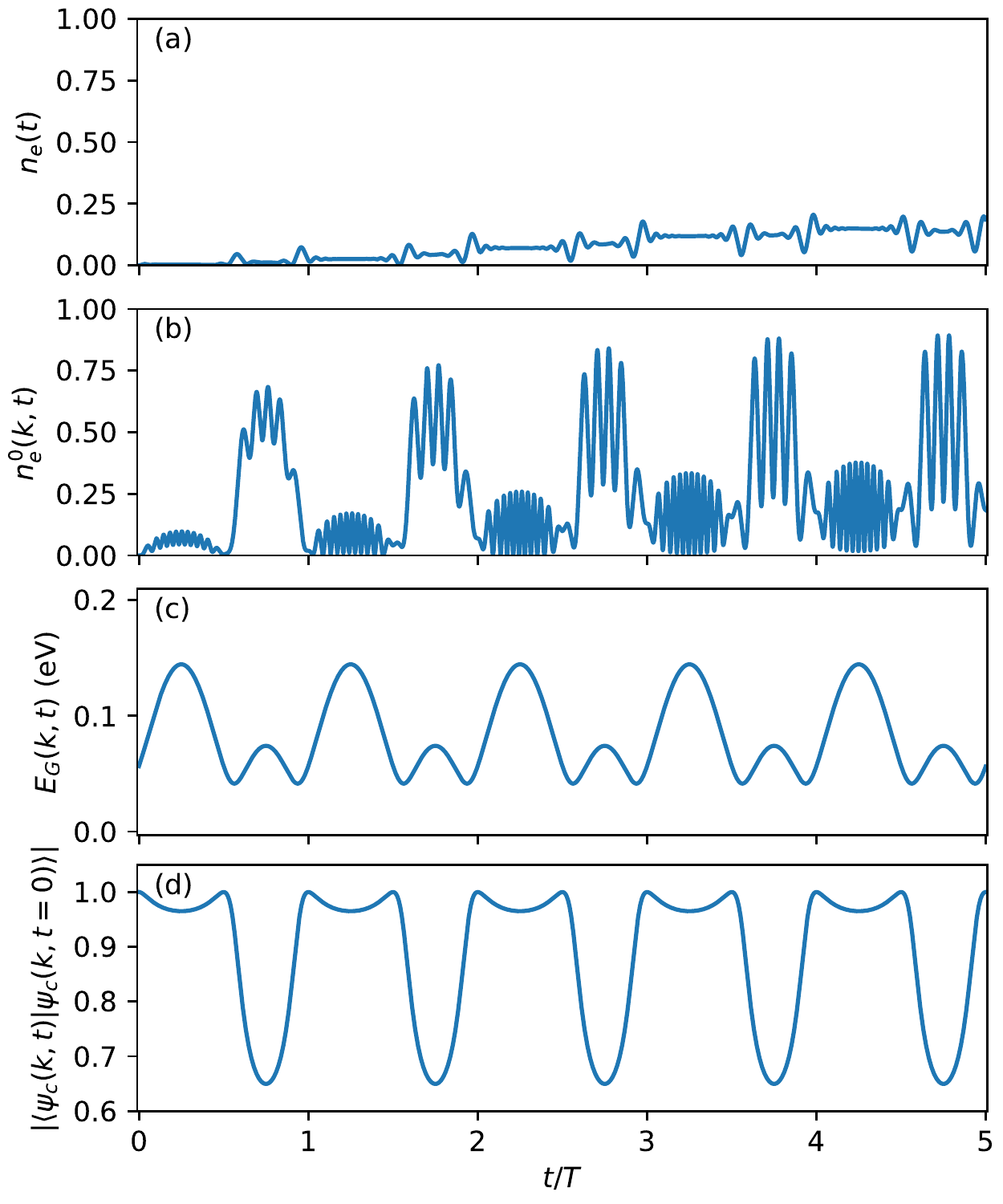}
	\caption{
	\textbf{Comparison of excited state population dynamics between two definitions at $k=0.033\pi$.} (a) Time evolution of the excited state population $n_e (k, t)$ defined with respect to the adiabatic conduction band for a periodic modulation $\mu(t)/\nu \in [-2.02, -1.92]$ starting at $\mu(t_0) = \mu_0$ for five full cycles. (b) The same as (a) but for $n^0_e (k, t)$ defined with respect to the initial conduction band. (c) Time evolution of the energy gap $E_G(k, t)$. (d) The absolute value of the overlap between adiabatic conduction band wavefunction $\psi_c(k, t)$ and the initial conduction band wavefunction $\psi_c(k, t=0)$ as a function of $t/T$. 
	}
	\label{fig: nek0}
\end{figure*}

For the carrier density analysis of the time-dependent BdG model in the main text, we define the excited state population $n_e(k, t) = \abs{\ov{\psi(k, t)}{\psi_c(k, t)}}^2$ with respect to the adiabatic conduction band. Here we investigate an alternative definition of $n^0_e(k, t) = \abs{\ov{\psi(k, t)}{\psi_c(k, t=0)}}^2$, which is defined with respect to the conduction band of the initial unperturbed Hamiltonian. 

In Fig.~\ref{fig: nek0} (a,b), we plot the time evolution of $n_e(k, t)$ and $n^0_e(k, t)$ at $k=0.033\pi$ starting at $\mu(t_0) = \mu_0$ for five full cycles. The major variation of $n_e(k, t)$ occurs reasonably when the system approaches the minimum of the band gap $E_G(k, t)$, as shown in Fig.~\ref{fig: nek0} (c). In contrast, the curve of $n^0_e(k, t)$ shows strong oscillation even away from band gap minimum, which implies $n^0_e(k, t)$ is not a physically reasonable definition in practice. Nevertheless, we observe that if one discards the fast oscillation, $n^0_e(k, t)$ reaches a value of about $0.2$, in a fairly good agreement with $n_e(k, t)$. This can be understood by checking the overlap between two reference conduction band wavefunctions $\abs{\ov{\psi_c(k, t)}{\psi_c(k, t=0)}}$, as plotted in Fig.~\ref{fig: nek0} (d). $\psi_c(k, t)$ agrees with $\psi_c(k, t=0)$ at $t=\frac{i}{2}T$ with $i=0, \dots 10$ where the perturbation in $\mu(t)$ vanishes. $\psi_c(k, t)$ deviates more from $\psi_c(k, t=0)$ when the perturbation in $\mu(t)$ increases, and the deviation grows more rapidly when the perturbation reduces the band gap. 

\section{Significance of gauge and phase corrections for the \textit{ab initio} quantum dynamics simulations}

\begin{figure*}
	\centering
 \includegraphics[width=\textwidth]{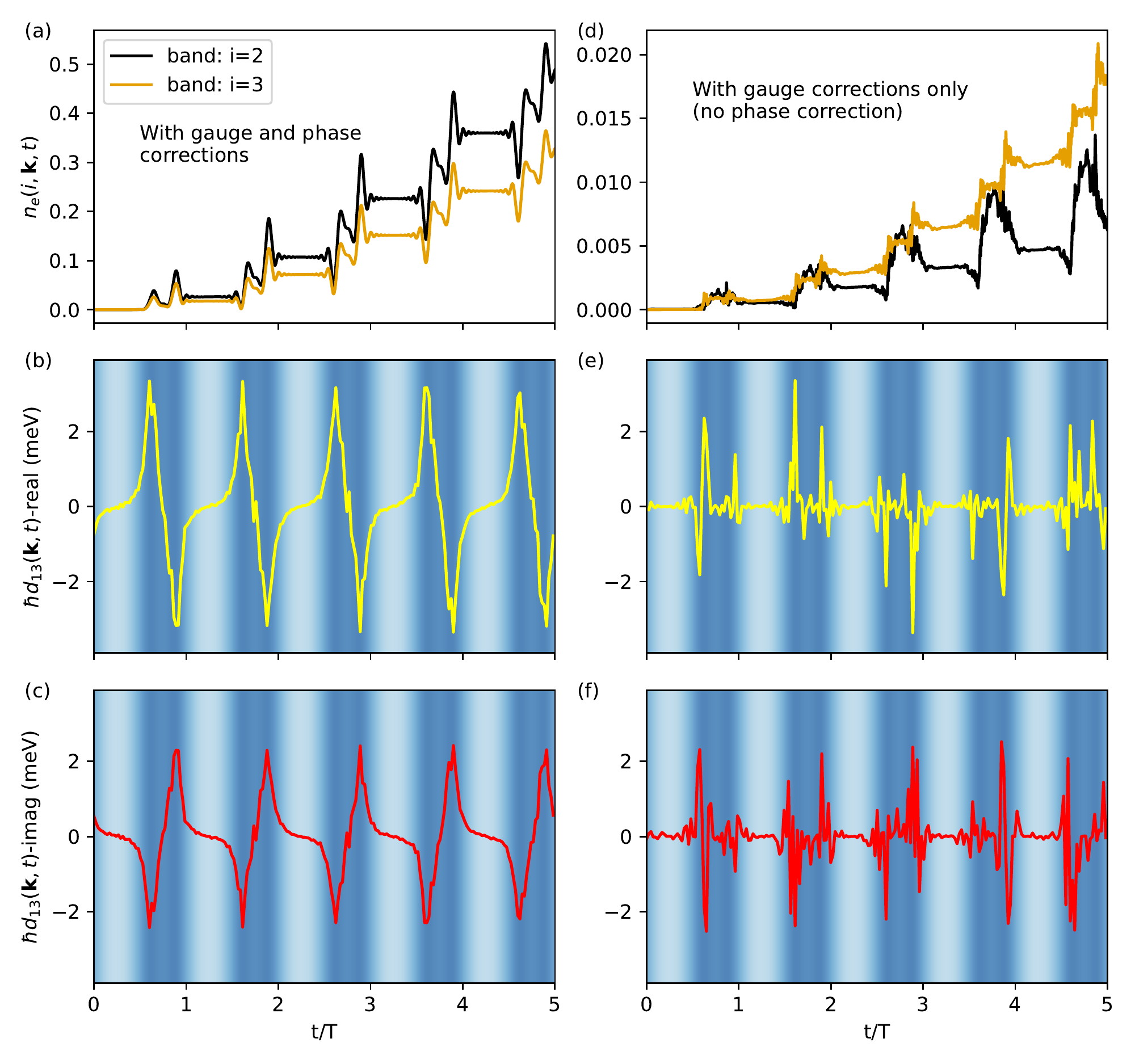}
	\caption{
	\textbf{Importance of phase correction on the simulation results.} (a) Occupancy of the two lowest-energy conduction bands, $n_e(i, \bk, t)$ with $i=2,3$, as a function of simulation time $t$ for $5$ phonon periods starting with $t_0=0$ (the equilibrium configuration with zero phonon displacement) at $\bk=\mathbf{b}_1/720 + \mathbf{b}_2/720 + \mathbf{b}_3/80$. Here $(\mathbf{b}_1, \mathbf{b}_2, \mathbf{b}_3)$ are the reciprocal primitive vectors. (b, c) The real and imaginary part of the time-dependent complex NAC amplitude $\hbar d_{13} (\bk, t)$ between the top valence band ($i=1$) and lowest conduction band ($i=3$) at the same $\bk$-point as (a), with real part in yellow and imaginary part in red. The background dark blue to light blue indicates a band gap, $E_\text{G} = \varepsilon_2 (\bk, t) - \varepsilon_1 (\bk, t)$, varying from $27$ meV to $132$ meV. (d, e, f) Similar results to (a, b, c) with the exception that the phase correction is not applied to the calculation. (a, b, c) is the reproduction of fig.3(a, b) to facilitate the comparison. To clearly show the random jumps in the real and imaginary part of NAC due to the uncorrected random phases in the wavefunctions, we plot the curves in (b,c,e,f) for every $20$ points of the time mesh. 
	}
	\label{fig: phase}
\end{figure*}

In the main text, we have discussed the $S_z$ and overlap gauge corrections and phase correction for the wavefunctions, which are crucial to produce physically reasonable quantum dynamics simulation results by fixing the artificial random jumps in the otherwise uncorrected time-dependent NAC amplitudes. For a specific example, we compare the simulation results with and without phase correction in Fig.~\ref{fig: phase} (a) and (c). The variation of the conduction band occupations $n_e(i\in[2,3], \bk, t)$ are found to be suppressed by over one order of magnitude without applying the phase correction. This can be understood by examining the effect of phase correction on the complex NAC amplitudes, as contrasted in Fig.~\ref{fig: phase} (b, c, e, f). While the real and imaginary part of the time-dependent NAC amplitude is quite smooth with phase correction, they show artificial random jumps without the phase correction. Similar random jumps can be observed without gauge corrections.

\end{document}